\documentclass[runningheads]{llncs}
\usepackage{graphicx}

\usepackage{graphicx}
\usepackage{amsfonts}
\usepackage{amsmath}
\usepackage{booktabs}
\usepackage{adjustbox}
\usepackage{subfig}
\usepackage{multirow}
\usepackage{multicol}
\begin{document}
\title{Dynamic GNNs for Precise Seizure Detection and Classification from EEG Data}
\titlerunning{Dynamic Multi-Context GNN for Seizure Detection \& Classification }
%
%

\author{Arash Hajisafi\inst{1}\orcidID{0009-0000-1056-0640}\and
Haowen Lin\inst{1}\orcidID{0000-0002-4097-1907}\and
Yao-Yi Chiang\inst{2}\orcidID{0000-0002-8923-0130} \and Cyrus Shahabi\inst{1}\orcidID{0000-0001-9118-0681}}

\authorrunning{Hajisafi et al.}
%

\institute{University of Southern California, Los Angeles, CA, USA\\
\email{\{hajisafi,haowenli,shahabi\}@usc.edu}
\and
University of Minnesota, Minneapolis, MN, USA
\email{yaoyi@umn.edu}}

\maketitle              
\begin{abstract}

Diagnosing epilepsy requires accurate seizure detection and classification, but traditional manual EEG signal analysis is resource-intensive. Meanwhile, automated algorithms often overlook EEG's geometric and semantic properties critical for interpreting brain activity. This paper introduces NeuroGNN, a dynamic Graph Neural Network (GNN) framework that captures the dynamic interplay between the EEG electrode locations and the semantics of their corresponding brain regions. The specific brain region where an electrode is placed critically shapes the nature of captured EEG signals. Each brain region governs distinct cognitive functions, emotions, and sensory processing, influencing both the semantic and spatial relationships within the EEG data. Understanding and modeling these intricate brain relationships are essential for accurate and meaningful insights into brain activity. This is precisely where the proposed NeuroGNN framework excels by dynamically constructing a graph that encapsulates these evolving spatial, temporal, semantic, and taxonomic correlations to improve precision in seizure detection and classification. Our extensive experiments with real-world data demonstrate that NeuroGNN significantly outperforms existing state-of-the-art models.

\end{abstract}
\section{Introduction}
Seizure detection and classification from EEG data is crucial for diagnosing epilepsy, a neurological disorder affecting around 1\% of the global population, with classification being particularly important for precise treatment \cite{falco2020epilepsy}. The conventional approach to seizure detection and classification is largely manual, requiring skilled personnel to analyze extensive EEG recordings. This manual process is time-consuming, expensive, and prone to human errors \cite{jiang2017seizure,tang2022selfsupervised}. 

Automated approaches using Convolutional Neural Networks (CNNs)~\cite{shoeibi2021epileptic} and Recurrent Neural Networks (RNNs)~\cite{roy2018deep} have been explored to tackle these challenges. However, these architectures do not capture the relationships between EEG observations in various brain regions, a crucial aspect of understanding the underlying neurological dynamics during seizures. Graph Neural Networks (GNNs) can capture relationships among EEG data points using a graph representation, yet existing GNN-based approaches either rely on predefined relations to generate a static graph or only use temporal correlations for dynamic graph generation, overlooking spatial and semantic relationships \cite{hajisafi2023learning,tang2022selfsupervised}. 
Such approaches miss out on leveraging multiple informative contexts and often result in poor performance, especially in scenarios marked by data scarcity, such as in classifying rare seizure types. Toward this end, a comprehensive solution is needed to automatically capture both static and dynamic evolving relationships and incorporate temporal, spatial, and semantic contexts simultaneously. This would allow for a detailed representation of the underlying correlations within the brain, enabling an end-to-end methodology for seizure detection and classification while overcoming the limitations of training under data scarcity.


Drawing inspiration from BysGNN~\cite{hajisafi2023learning}, a GNN framework that creates a dynamic multi-context graph for forecasting visits to points of interest (POIs) based on spatial, temporal, semantic, and taxonomic contexts, we aim to generate a similar multi-context graph for our EEG data. However, adapting BysGNN as is to our domain presents several challenges. First, the shift from time-series forecasting in BysGNN to a multivariate time-series classification task in our work requires a change in architectural design. While BysGNN focuses on patterns within individual graph entities for node regression, our goal requires learning broader patterns across entities to perform graph classification. Second, EEG signals represent neurological events, which causes the translation of multi-context correlations from the POI domain to EEG data to be complex due to the underlying neurological mechanisms in brain activity and functional dependencies. Specifically, this translation is nontrivial because (1) the semantic contexts and relationships in brain areas are based on the functional roles of the corresponding part of the brain or broader brain region, which differs from the semantics of POI visits; (2) unlike POI's geographical coordinates, spatial contexts in EEG data are based on the brain's anatomy and EEG electrode placements; (3) the higher-level patterns in the brain should be defined based on the field of neuroscience, which is not as straightforward as defining high-level visit patterns to POIs. Finally, there are far fewer training samples for seizure classification, especially for rare seizure types, which necessitates designing a pretraining strategy aligned with the multi-context correlation notion to achieve optimal seizure detection and classification performance.

To address these challenges, we propose NeuroGNN, a GNN-based framework designed to dynamically construct a graph encapsulating multi-context correlations in EEG data. The correlations are characterized by spatial proximity of electrode placements on the scalp, temporal dependencies within and between time series from different electrodes, semantic similarities derived from neurological brain functions of respective electrode placements, and taxonomic correlations across broader brain regions. The broader regions are defined through neuroscientific meta-nodes generated from EEG electrodes, extending the multi-context definitions to these meta-nodes. Our new design incorporates hierarchical pooling to handle node and meta-node representations and employs Bidirectional Gated Recurrent Units (BiGRUs) to enhance the intra-series capture and characterization of time-series data. To address the challenge of small training samples and achieve optimal performance across both detection and classification, we develop a new pretraining strategy with a learning objective that aligns with our graph representation, encompassing both nodes and meta-nodes. Learning to generate this multi-context graph representation allows NeuroGNN to outperform state-of-the-art methods for seizure detection and classification in real-world EEG diagnosis scenarios. Through comprehensive ablation studies, we demonstrate that each multi-context correlation modality significantly contributes to the overall performance enhancement. This is particularly noticeable in rare classes with training under data scarcity.



\section{Related Work}
\label{sec:related-works}
EEG signals are commonly used to detect epileptic seizures (e.g., \cite{shoeibi2021epileptic}). Since manual EEG analysis by trained physicians is extremely resource and time-intensive, much recent research has focused on using machine learning approaches to automate the diagnostic process. Diagnosing epileptic seizures can be divided into seizure detection and seizure classification. Seizure detection aims to detect abnormal patterns in EEG data. Initially, for seizure detection, researchers started with conventional machine learning techniques, where they first extracted hand-engineered frequency features from the recordings and applied traditional methods such as K-Nearest Neighbors and random forests to detect abnormal EEG signals \cite{lopez2015automated}. Shortly afterward, Roy et al. suggested employing time-distributed neural networks, such as recurrent neural networks, to learn directly from the data without any explicit pre-processing step~\cite{roy2018deep}. Sharathappriyaa et al. propose to compress EEG signals into a latent space representation using Autoencoders and then detect epileptic seizures \cite{sharathappriyaa2018auto}.  

Seizure classification differentiates between seizure types like focal or generalized seizures. Most of the works in seizure classification adopt CNNs for EEG-based seizure prediction \cite{shoeibi2021epileptic}. For example, Bhattacharyya et al. first transform the 1D signal of each EEG channel into 2D using wavelet transform and then apply a CNN to detect and classify seizure subtypes \cite{bhattacharyya2017tunable}. However, CNN-based methods often overlook the complex spatial structures of EEG sensors and brain geometry \cite{tang2022selfsupervised}. To address this, recent efforts have explored graph-based neural networks, which better capture these spatial relationships, achieving enhanced detection and classification results~\cite{tang2022selfsupervised}. Yet, existing GNN studies on EEG data present a significant limitation: they either assume a static graph structure based solely on EEG sensor distances or generate dynamic graphs considering only the temporal relationships between EEG. Both approaches are constrained, representing only a single type of relationship in the graph structure, spatial or temporal, and thus fail to fully capture the complex interactions among EEG sensors and their signals. Our model addresses this gap by effectively fusing complex relationships from multiple dimensions to create dynamic graphs that improve automatic seizure detection and classification.

\begin{figure*}[t]
  \centering
          \begin{adjustbox}{width=\linewidth, center}
\includegraphics{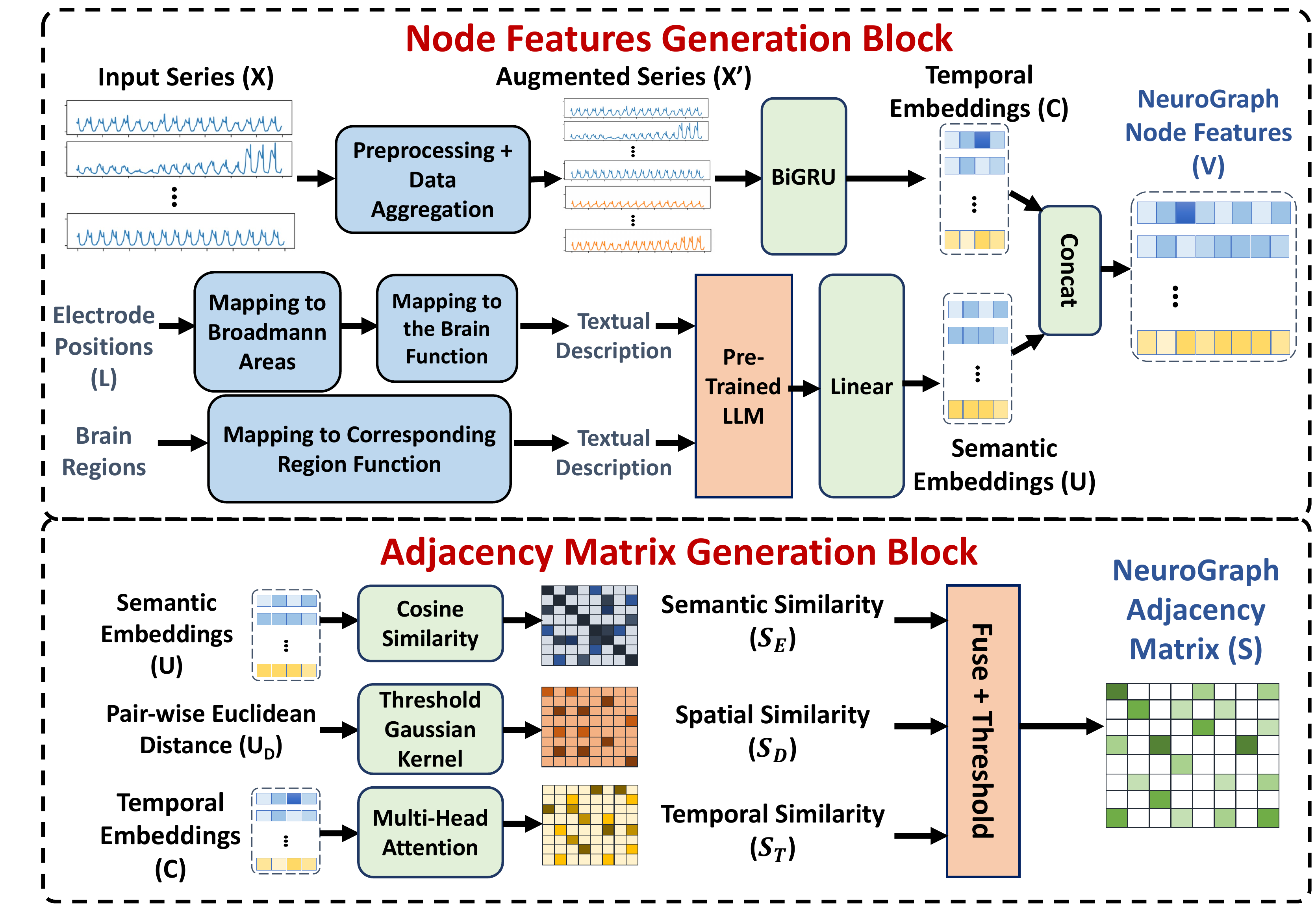}
        \end{adjustbox}
        \vspace{-18pt}
  \caption{\textbf{NeuroGNN Graph Construction}}
  \label{fig:neurognn-graph-construction}
  \vspace{-8pt}
\end{figure*}

\begin{figure*}[t]
  \centering
          \begin{adjustbox}{width=\linewidth, center}
\includegraphics{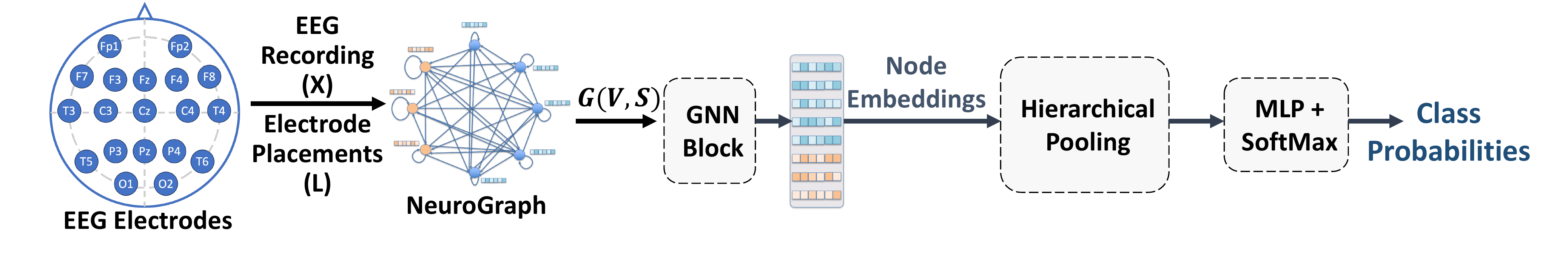}
        \end{adjustbox}
        \vspace{-20pt}
  \caption{\textbf{Seizure Detection and Classification using NeuroGraph}}
  \vspace{-15pt}
  \label{fig:seizure-classification}
\end{figure*}

\section{Methodology: NeuroGNN}
\label{sec:method}
Given EEG recordings within a time-series window $X$ and the 3D coordinates $L$ of electrode placements on the scalp, we first construct the dynamic NeuroGraph $G$, integrating multi-context relations, as illustrated in Figure~\ref{fig:neurognn-graph-construction}. This graph's nodes encapsulate current temporal EEG data and corresponding brain semantics. Edges fuse spatial, temporal, and semantic information to model complex inter-node relationships. Subsequently, as depicted in Figure~\ref{fig:seizure-classification}, the GNN block updates the node embeddings, leveraging these multi-contextual edges. Hierarchical pooling aggregates these embeddings into a single graph-level vector, which is then classified via an MLP, enabling both binary classification for seizure detection and multi-label classification for seizure type prediction. Detailed procedures are in subsequent sections.

\vspace{-10pt}
\subsection{Node Features Generation}
\vspace{-5pt}
The process begins by preprocessing the time-series data and generating new aggregated series to form six meta-nodes, each representing a distinct brain region as identified in \cite{ackerman1992discovering}. These meta-nodes, representing regional neural activities, complement EEG nodes that capture only localized activities, facilitating a multi-scale representation for learning taxonomic correlations. Meta-nodes aggregate EEG data within their regions through an averaging function 
on the input \(\mathbf{X} \in \mathbb{R}^{N \times T}\), producing \(X' \in \mathbb{R}^{N' \times T}\) where \(N'=N+6\) (representing the addition of six brain regions).

Next, to capture intra-series temporal correlations, \(X'\) is fed into a Bidirectional Gated Recurrent Unit (BiGRU) \cite{cho2014learning}. BiGRU's bidirectional processing provides a comprehensive representation of the given time-series window, which is crucial for seizure detection and classification. The BiGRU generates hidden states for each node, denoted as \(h_{f,i} \in \mathbb{R}^{M}\), representing forward temporal dynamics and \(h_{b,i} \in \mathbb{R}^{M}\), representing backward temporal dynamics from the forward and backward passes, respectively, where $M$ shows the hidden dimension. The final node embedding is obtained by concatenating the two hidden states, \(c_i = (h_{f,i} \mathbin\Vert h_{b,i})\), forming an embedding matrix \(C \in \mathbb{R}^{N' \times 2M}\).

In parallel, NeuroGNN captures semantic correlations among EEG nodes and meta-nodes using a pre-trained Large Language Model (LLM). The semantic foundations are based on Brodmann areas, which segment the cerebral cortex into distinct functional regions \cite{zilles2018brodmann}. A mapping between EEG electrode placements $L$ and Brodmann areas is established as per \cite{scrivener2022variability}. Textual descriptions encapsulating the functional roles of the corresponding Broadmann areas and brain regions are generated for each EEG node and meta-node based on reputable scientific resources~\cite{strotzer2009one,zilles2018brodmann}. These descriptions are tokenized and fed into a pre-trained MPNet language model~\cite{song2020mpnet}, which is fine-tuned via a linear layer during training to optimize semantic embeddings for the downstream task, resulting in a semantic embeddings matrix \(U \in \mathbb{R}^{N' \times K}\), where \(K\) denotes the semantic embedding dimensionality.

The final feature vector for each node and meta-node is obtained by concatenating the temporal and semantic embeddings, yielding the NeuroGraph node features matrix \(V \in \mathbb{R}^{N' \times (2M+K)}\).

\vspace{-15pt}
\subsection{Adjacency Matrix Generation}
\vspace{-5pt}
Next, we construct NeuroGraph's adjacency matrix to dynamically capture multi-context correlations among EEG and meta-nodes.

First, we compute the semantic similarity matrix \(S_E\) using cosine distance on semantic embeddings from the node features generation layer. In parallel, we compute spatial and temporal similarities. Spatial similarities are derived by first calculating Euclidean distances among EEG electrodes for EEG node pairs, and average distance for meta-node pairs and meta-node to EEG node pairs. The spatial similarity matrix \(S_D\) is then formed by applying a Gaussian kernel \cite{shuman2013emerging} to convert distance metrics into similarity scores, controlled by a threshold~\(\tau\). Temporal similarities are captured by passing temporal embeddings \(C\) to a Multi-Head Attention layer \cite{vaswani2017attention}, resulting in a temporal similarity matrix~\(S_T\).

To fuse these similarities, we employ a gating mechanism, \(S_{Gate} = (1-\alpha)S_E + \alpha S_D\) to control influence of temporal similarities, forming an un-thresholded adjacency matrix \(S' = S_{Gate} \odot S_T\), with \(\alpha\) as a learnable parameter between 0 and 1 to adjust the balance between the importance of spatial and semantic similarities. A thresholding step \cite{hajisafi2023learning} refines \(S'\) to obtain the final adjacency matrix \(S\), retaining reliable similarity scores and discarding noisy ones.

\vspace{-10pt}
 \subsection{Prediction using the Generated NeuroGraph}
With the generated node features \(V\) and adjacency matrix \(S\), the NeuroGraph \(G=(V,S)\) for a given window is formed, encapsulating multi-context relationships among EEG nodes and meta-nodes. NeuroGraph is then passed through a GNN block employing a modified Graph Convolutional Networks (GCN) variant~\cite{kipf2017semisupervised}, omitting the normalization term and adding residual connections between the message passing layers, to preserve directed relationships in NeuroGraph and mitigate oversmoothing~\cite{chen2020measuring}, respectively. This yields the node embeddings matrix \(V' \in \mathbb{R}^{N' \times Z}\), where \(Z\) represents the embedding dimension.

Next, a hierarchical pooling mechanism aggregates refined node representations into a single graph embedding vector. Initially, within each brain region, a max-pooling operation aggregates EEG node embeddings. Similarly, a max-pooling operation across all meta-nodes constructs a single embedding vector for meta-nodes. These pooled embeddings are concatenated, forming a collective representation of region-wise and meta-wise node information. A feature-wise mean-pooling operation on this concatenated representation derives the graph embedding vector \(\mathbf{g} \in \mathbb{R}^{Z}\), encapsulating critical graph information.

Finally, vector \(\mathbf{g}\) is processed through a Multi-Layer Perceptron (MLP) with a Softmax activation to obtain the final class probabilities. For training detection and classification models, binary and multi-class cross-entropy are utilized as loss functions, respectively.

\vspace{-7pt}
\subsection{Pretraining}
\label{method:pretraining}
Pretraining can mitigate issues related to the scarcity of labeled data and out-of-distribution predictions in many scientific ML applications \cite{zeiler2014visualizing,hendrycks2019using}. Therefore, we assess NeuroGNN's performance with weights initialized after a pretraining phase on a self-supervised task. The chosen pretraining task is a node-level regression task, which forecasts future values for the upcoming 12 seconds from a 60s preprocessed EEG clip for each EEG node and meta-node. We design two training objectives;
{\bf Primary Objective:} Minimizing Forecasting Mean Squared Error (MSE) Loss across all EEG nodes and meta-nodes, and {\bf Secondary Objective:} Ensuring Consistency Loss to maintain forecast consistency between individual nodes and their corresponding meta-nodes. The overall training objective combines both objectives with a weighted sum.


\section{Experiments}
\label{sec:experiments}
\subsection{Experimental Setup}
\subsubsection{Dataset and Data Splits}
We conducted our experiments using the Temple University Hospital EEG Seizure Corpus (TUSZ) v1.5.2 \cite{shah2018temple,obeid2016temple}. The dataset comprises $3,050$ annotated seizure events from over $300$ patients across eight seizure types, with data recorded using 19 electrodes from the standard 10-20 EEG system. We follow \cite{tang2022selfsupervised} to preprocess and group the seizures into four classes.
Table \ref{tabl:dataset-stats} details the statistics of the data used in our experiments.
The training set was randomly split into training and validation sets with a 9:1 ratio.


\vspace{-17pt}
\subsubsection{\indent Preprocessing}
Following prior methodologies \cite{tang2022selfsupervised,ahmedt2020neural,asif2020seizurenet}, EEG recordings were resampled to 200Hz and divided into 60-second non-overlapping windows (clips). For seizure detection, all EEG clips are utilized; a clip is labeled as a seizure if it contains at least one type of seizure. For seizure classification, only clips with a single seizure type are used. If a seizure ends and another begins within a clip, it is truncated and zero-padded to maintain a 60-second length. Each 60-second clip is further divided into 1-second segments, with the Fast Fourier Transform (FFT) applied to each segment to obtain log amplitudes of non-negative frequency components, as per \cite{tang2022selfsupervised}. Each 60-second clip, now a sequence of 60 log-amplitude representations, is used for classification.


\begin{table*}[t]
\centering
\caption{Summary of TUSZ v1.5.2 used in our study}
\label{tabl:dataset-stats}
\begin{adjustbox}{width=\linewidth, center}
\begin{tabular}{|l|c|c|c|c|c|c|c|c|c|c|c|c|c|c|}
\hline
\multirow{2}{*}{\fontsize{10}{12}\selectfont Set} & \multicolumn{2}{c|}{\fontsize{10}{12}\selectfont EEG Files} & \multicolumn{2}{c|}{\fontsize{10}{12}\selectfont Total Seizure Duration} & \multicolumn{2}{c|}{\fontsize{10}{12}\selectfont Patients} & \multicolumn{2}{c|}{\fontsize{10}{12}\selectfont CF Seizures} & \multicolumn{2}{c|}{\fontsize{10}{12}\selectfont GN Seizures} & \multicolumn{2}{c|}{\fontsize{10}{12}\selectfont AB Seizures} & \multicolumn{2}{c|}{\fontsize{10}{12}\selectfont CT Seizures} \\ \cline{2-15} 
 & \fontsize{10}{12}\selectfont Non-Seizure & \fontsize{10}{12}\selectfont Seizure & \fontsize{10}{12}\selectfont Non-Seizure & \fontsize{10}{12}\selectfont Seizure & \fontsize{10}{12}\selectfont Non-Seizure & \fontsize{10}{12}\selectfont Seizure & \fontsize{10}{12}\selectfont Seizures & \fontsize{10}{12}\selectfont Patients & \fontsize{10}{12}\selectfont Seizures & \fontsize{10}{12}\selectfont Patients & \fontsize{10}{12}\selectfont Seizures & \fontsize{10}{12}\selectfont Patients & \fontsize{10}{12}\selectfont Seizures & \fontsize{10}{12}\selectfont Patients \\ \hline
Train Set & 3730 & 869 & 705h 29m & 47h 26m & 390 & 202 & 1,868 & 148 & 409 & 68 & 50 & 7 & 48 & 11 \\ \hline
Test Set & 670 & 230 & 135h 46m & 14h 45m & 10 & 35 & 297 & 24 & 114 & 11 & 49 & 5 & 61 & 4 \\ \hline
\end{tabular}
\end{adjustbox}
\vspace{-20pt}
\end{table*}




\vspace{-15pt}
\subsubsection{\indent Training} 
Training was conducted on a RTX 3090 GPU with 24GB of memory under an Ubuntu 20.04 setup with CUDA version 11.4, and PyTorch v1.13.0. Without pretraining, model weights were initialized using the Xavier initialization method \cite{glorot2010understanding}. The Adam optimizer \cite{kingma2014adam} was utilized for training, paired with a cosine annealing learning rate scheduler \cite{loshchilov2016sgdr}. The training was continued until the evaluation loss increased in 5 consecutive epochs or reached a maximum of 100 epochs. $L2$ regularization was applied to model weights to prevent overfitting. In seizure detection and classification, the model checkpoint achieving the best Area Under the Receiver Operating Characteristic (AUROC) and weighted F1 score on the validation set across different epochs were preserved, respectively.
The pretraining task also ran for 300 epochs. The obtained weights for Node Features Generation, Adjacency Matrix Generation, and GNN Block parameters then initialized the model for the following downstream task.

\vspace{-15pt}
\subsubsection{\indent Baselines}

We benchmarked NeuroGNN against: (1) \textbf{LSTM} \cite{hochreiter1997long}, for encoding sequential data; (2) \textbf{Dense-CNN} \cite{saab2020weak} that used a densely connected architecture for seizure detection; (3) \textbf{CNN-LSTM} \cite{ahmedt2020neural} integrated 2D convolutions with LSTM for seizure classification; (4) \textbf{Corr-DCRNN} \cite{tang2022selfsupervised}, a GNN-based model that utilized the DCRNN \cite{li2018diffusion} framework on graphs formed from EEG data cross-correlation at each timestep; and (5) \textbf{Dist-DCRNN} \cite{tang2022selfsupervised}, another GNN variant leveraging DCRNN on a graph constructed from Euclidean distances of EEG electrode placements. The latter two represent current state-of-the-art (SOTA) in this domain.

\vspace{-15pt}
\subsubsection{\indent Hyperparameter Setting}
Cross-validation was used to select hyperparameters for optimal performance. Key settings include an initial learning rate of $0.0002$, batch size of $40$, hidden BiGRU dimension $M$ and semantics embedding dimension $K$ both at $512$. Multi-Head Attention utilized $8$ heads, while graph convolution node embedding dimension $Z$ was set to $256$. Gaussian kernel threshold $\tau$ was set to twice the standard deviation of EEG electrode distances. For pretraining, the MSE loss objective weight was set to 0.9, and the Consistency loss objective weight was adjusted to 0.1. These settings were consistent across both detection and classification tasks.

\begin{table*}[t]
\caption{Experiment results. The highest and second highest scores are denoted in bold and italics with underline, respectively. ``Improvement'' shows the percentage improvement of NeuroGNN compared to the best-performing baseline. The mean and standard deviations are from three random runs.}
\label{tab:performance-results}
    \centering
\resizebox{0.9\textwidth}{!}{%
    \begin{tabular}{|c|c|c|c|c|}
    \toprule
        \bf{Method} & 
        \multicolumn{2}{c|}{\bf{Seizure Detection (AUROC)}} & 
        \multicolumn{2}{c|}{\bf{Seizure Classification (Weighted F1)}} \\ 
        \cline{2-5}
        & Without Pretraining & With Pretraining & Without Pretraining & With Pretraining \\
        \hline
        LSTM & $0.715 \pm 0.016$ & -- & $0.686 \pm 0.020$ & -- \\ 
        Dense-CNN & $0.796 \pm 0.014$ & -- & $0.626 \pm 0.073$ & -- \\ 
        CNN-LSTM & $0.682 \pm 0.003$ & -- & $0.641 \pm 0.019$ & -- \\ 
        \hline
        Corr-DCRNN & \underline{\emph{0.804} $\pm$ \emph{0.015}} & $0.850 \pm 0.014$ & \underline{\emph{0.701} $\pm$ \emph{0.030}} & \underline{\emph{0.749} $\pm$ \emph{0.017}} \\ 
        Dist-DCRNN & $0.793 \pm 0.022$ & \underline{\emph{0.875} $\pm$ \emph{0.016}} & $0.690 \pm 0.035$ & \underline{\emph{0.749} $\pm$ \emph{0.028}} \vspace{1pt}\\ 
        \hline
        NeuroGNN & \bf{0.847} $\pm$ \bf{0.007} & \bf{0.876} $\pm$ \bf{0.002} & \bf{0.790} $\pm$ \bf{0.006} & \bf{0.792} $\pm$ \bf{0.004} \\
        \hline
        Improvement & +5.35\% & +0.11\% & +12.7\% & +5.74\% \\
        \bottomrule
    \end{tabular}
}
\vspace{-20pt}
\end{table*}



\vspace{-10pt}
\subsection{Results}

\subsubsection{NeuroGNN Performance}
Table~\ref{tab:performance-results} presents the results using the evaluation metrics following convention~\cite{tang2022selfsupervised,asif2020seizurenet,ma2023tsd,ahmedt2020neural}. NeuroGNN outperforms baseline models in seizure detection and classification in both pretraining scenarios. In seizure detection, non-pretrained Corr-DCRNN shows a slight 1.01\% improvement over Dense-CNN, while NeuroGNN boosts the AUROC score by over 5\% compared to SOTA Corr-DCRNN, underscoring the benefits of capturing multi-context correlations. Both Dist-DCRNN and Corr-DCRNN, tackling different correlation aspects, spatial and temporal, individually, yield similar performance on both tasks. Conversely, NeuroGNN, integrating multi-context correlations within a unified framework, shows substantial performance improvement against DCRNN-based baselines. Particularly in classification, NeuroGNN significantly improves the weighted F1 score of DCRNN-based baselines by 12.7\% without pretraining and by 5.74\% after pretraining, respectively.

The improvement in seizure detection achieved by NeuroGNN over baselines, while noteworthy, is less significant compared to the enhancement in seizure classification performance. This difference can be attributed to the binary nature of the detection task and the abundance of training data, rendering it a less complex task. 
Even a simpler model like Dense-CNN holds up well against DCRNN-based models without pretraining. Conversely, the classification task, a 4-class problem with imbalanced class distributions (Table \ref{tabl:dataset-stats}), poses more challenges. The limited training samples for certain seizure types highlight the importance of multi-context correlations, illustrated by over 12\% improvement in weighted F1 score by NeuroGNN without pretraining.
\vspace{-15pt}
\subsubsection{Impact of Pretraining}
Consistent with findings in \cite{tang2022selfsupervised}, our results validate the positive effect of pretraining on performance. Our specific pretraining task, aligned with NeuroGNN's design involving EEG nodes and meta-nodes representing brain regions, enhances performance, leading to a 3.42\% boost in the AUROC metric for seizure detection. However, the increase in the weighted F1 score for classification is minor. This indicates that while pretraining aids in mitigating data scarcity and out-of-distribution challenges, the multi-context correlations captured by NeuroGNN intrinsically address these issues, making the benefits of pretraining less significant for NeuroGNN compared to DCRNN-based models. This reduction in pretraining necessity also implies a saving in computational resources and time, positioning NeuroGNN as a more resource-efficient and practical option for real-world applications.

\begin{table*}[t!]
\caption{Ablation study results, showing median performance over three trials and percentage changes relative to NeuroGNN. Bold and underline percentages highlight the largest and second-largest changes per metric.}
\label{tab:abl-study}
    \centering
\resizebox{0.8\textwidth}{!}{%
    \begin{tabular}{|c||c|c|c|c|c|c|}
    \toprule
        ~ & NeuroGNN & w/o. TemporalCorr & w/o. Semantics & w/o. Space & w/o. Meta-Nodes \\
        \hline
        Detection & \textbf{0.847} & 0.831 & 0.828 & 0.818 & 0.829\\
         (AUROC) & -- & -2.12\% & \underline{\emph{-2.47}}\% & \textbf{-3.65}\% &  -2.36\%\\
        \hline
        Classification & \textbf{0.790} & 0.673 & 0.719 & 0.727 & 0.714\\
        (Weighted F1) & -- & \textbf{-14.81\%} & -8.99\% & -7.97\% & -\underline{\emph{9.62}}\%\\

        \bottomrule
    \end{tabular}
    }
    \vspace{-18pt}
\end{table*}

\vspace{-13pt}
\subsection{Ablation Study}
\vspace{-5pt}
In this section, we explore the impact of NeuroGNN's individual components on performance by creating four distinct variants, each removing one aspect of the multi-context correlations. The variants are:
(1) \textbf{W/O TemporalCorr:} NeuroGNN without temporal correlations for graph edge formation. (2) \textbf{W/O Semantics:} NeuroGNN without semantics in node features and graph edge weights. (3) \textbf{W/O Space:} NeuroGNN without spatial proximity information between nodes for graph edge formation. (4) \textbf{W/O Meta-Nodes:} NeuroGNN without aggregated time-series nodes representing brain regions.


We evaluated each variant on both seizure detection and classification tasks, with results presented in Table \ref{tab:abl-study}. As observed from the table, every component improves performance on both tasks. The ablation study shows a noticeable decrease in performance for the classification task in all configurations compared to the detection task. On average, the classification task experiences a 7.7\% larger drop in performance in all configurations. This observation supports our claim that multi-context correlations significantly ease the challenges posed by data scarcity in rare classes and the complexity of multi-class classification. For example, the classification task has significantly fewer samples than the detection task (3,730 non-seizure and 869 seizure events), particularly for the AB and CT classes, which have only 50 and 48 training events, respectively.

The analysis indicates that spatial correlations have the most significant impact on the detection task, while temporal correlations affect classification performance the most, leading to a 3.65\% and a substantial 14.81\% drop in performance, respectively. This can be explained by the nature of the tasks. In seizure detection, the goal is to pinpoint the occurrence of seizures, which often relate to unusual frequency patterns within a close neighborhood of EEG electrodes. On the other hand, classification requires a more detailed examination of frequency anomalies to distinguish the type of seizure, highlighting the importance of temporal relationships among different EEG electrodes and brain regions in associating an EEG clip with a specific seizure class.


\begin{figure*}[tbp]
    \centering
    \subfloat[Graph Embeddings from Dist-DCRNN]{%
        \includegraphics[width=0.49\textwidth]{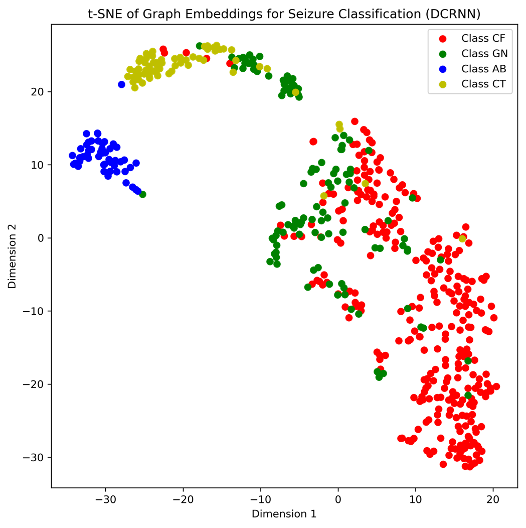}%
        \label{fig:dcrnn-graph-embs}%
    }\hfill
    \subfloat[Graph Embeddings from NeuroGNN]{%
        \includegraphics[width=0.49\textwidth]{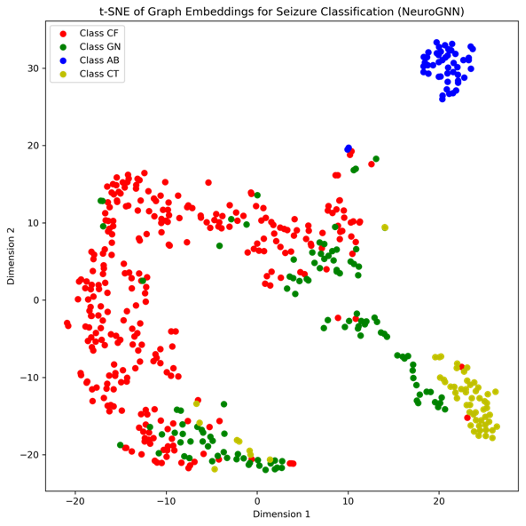}%
        \label{fig:neurognn-graph-embs}%
    }\hfill
    \caption{Visualization of graph embeddings for seizure classification test samples using trained Dist-DCRNN and NeuroGNN models. The colors represent the ground truth seizure labels.}
    \label{fig:graph-embeddings}
    \vspace{-17pt}
\end{figure*}

\vspace{-10pt}
\subsection{Analysis of Graph Embeddings}
\vspace{-5pt}
This section explores the effectiveness of graph embeddings generated by pretrained Dist-DCRNN and NeuroGNN on test samples for the classification task. The aim is to assess the effectiveness of the learned graph embeddings by NeuroGNN in comparison with the previous SOTA model in distinguishing different seizure types.  After training on seizure classification, both models are used in inference mode to obtain graph embeddings for test samples, with initial weights set based on pretraining for a fair comparison.

The graph embeddings for the seizure classification test samples are visualized in a 2D space using the t-SNE dimensionality reduction technique~\cite{van2008visualizing} as depicted in Figure~\ref{fig:graph-embeddings}. These figures reveal a notable pattern of interleaving among samples of the GN seizure class and samples from other classes, which corresponds to a lower True Positive Rate for the GN seizure class, indicating higher classification difficulty for this particular seizure type.

The quality of embeddings is assessed quantitatively by clustering the original high-dimensional graph embeddings into four clusters using the KMeans algorithm, reflecting the number of classes in the dataset. The average clustering purity metric is computed to measure the homogeneity of clusters regarding class membership, defined as the mean of the maximum class proportion across all clusters. The analysis shows that NeuroGNN achieves an average clustering purity value of $0.747$, compared to $0.717$ for Dist-DCRNN, representing an 11\% reduction in clustering impurity by NeuroGNN. This highlights NeuroGNN's superior ability to distinguish samples based on multi-context correlations.

\begin{table*}[t]
\centering
\caption{F1 Score Comparison between NeuroGNN and Dist-DCRNN for Seizure Classification Across Different Training Data Ratios. ``Improvement'' shows the percentage improvement of NeuroGNN over Dist-DCRNN.
}
\label{tab:performance-gap}
\begin{tabular}{|c|c|c|c|c|c|}
\hline
 & \multicolumn{5}{c|}{Training Data Ratio} \\
\hline
 & \hspace{10pt} 100\% \hspace{10pt} & \hspace{10pt} 80\% \hspace{10pt} & \hspace{10pt} 60\% \hspace{10pt} & \hspace{10pt} 40\% \hspace{10pt} & \hspace{10pt} 20\% \hspace{10pt} \\
\hline
Dist-DCRNN & 0.701 & 0.704 & 0.679 & 0.689 & 0.606 \\
\hline
NeuroGNN & 0.790 & 0.795 & 0.772 & 0.787 & 0.781 \\
\hline
Improvement & 12.7\% & 12.9\% & 13.7\% & 14.2\% & 28.9\% \\
\hline
\end{tabular}
\vspace{-17pt}
\end{table*}


\vspace{-12pt}
\subsection{Handling Scarce Training Data}
\vspace{-5pt}
Experiments were conducted to evaluate NeuroGNN and Dist-DCRNN's performance under different data availability scenarios by using subsampled versions of training data, while keeping the original class distribution through stratified sampling. The results, depicted in Table~\ref{tab:performance-gap}, show the median weighted F1 score in three independent runs for each sampling ratio in the training data, with evaluations performed on the entire test dataset to ensure consistency across all experiments.

The table illustrates that as the sampling ratio decreases, the performance gap between NeuroGNN and Dist-DCRNN increases, showcasing the robustness of NeuroGNN to smaller training sizes compared to Dist-DCRNN. Moreover, NeuroGNN's performance exhibits only slight degradation even with reduced training data, emphasizing the model's ability to effectively leverage multi-context correlations to address challenges posed by data scarcity, thereby maintaining a superior performance under data-constrained conditions.

\vspace{-9pt}
\section{Conclusion}
\label{sec:conclusion}
\vspace{-8pt}
We introduced NeuroGNN, a novel Graph Neural Network framework, to enhance seizure detection and classification from EEG data. While existing methods rely heavily on manual analysis or fail to capture the intricate brain dynamics, NeuroGNN dynamically integrates spatial, temporal, and semantic contexts. In our empirical evaluation, NeuroGNN consistently outperformed current state-of-the-art models, demonstrating its efficacy, especially when data is scarce for training, which is common when classifying rare seizures. 

\vspace{-6pt}
\section*{Acknowledgments}
Research supported by the National Science Foundation (NSF) under CNS-2125530 and the National Institute of Health (NIH) under grant 5R01LM014026. {\bf Disclaimer:} The views and conclusions contained herein are those of the authors and should not be interpreted as necessarily representing the official policies or endorsements, either expressed or implied, of NSF or NIH.

\vspace{-7pt}
\bibliographystyle{splncs04}

\bibliography{references}

\end{document}